\documentclass[12pt]{iopart}

\usepackage{epsf}  

\begin{document}

\jl{4}

\title{Strange Messages: Chemical and Thermal Freeze-out in 
       Nuclear Collisions}

\author{U Heinz}

\address{CERN/TH, CH-1211 Geneva 23, Switzerland; and Institut f\"ur \
         Theoretische Physik, Universit\"at Regensburg, 
         D-93040 Regensburg, Germany} 

\begin{abstract}
Thermal models are commonly used to interpret heavy-ion data on particle 
yields and spectra and to extract the conditions of chemical and thermal 
freeze-out in heavy-ion collisions. I discuss the usefulness and limitations 
of such thermal model analyses and review the experimental and theoretical 
evidence for thermalization in nuclear collisions. The crucial role of
correlating strangeness production data with single particle spectra and 
two-particle correlation measurements is pointed out. A consistent
dynamical picture for the heavy-ion data from the CERN SPS involves an
initial prehadronic stage with deconfined color and with an appreciable
isotropic pressure component. This requires an early onset of thermalization.
\end{abstract}

\pacs{24.10.Pa, 25.75.-q, 25.75.Dw, 25.75.Ld}



\section{Introduction}
\label{sec1}

The stated goal of the relativistic heavy-ion programs at CERN and BNL
is the study of the phase diagram of strongly interacting matter at
high temperatures and densities and the search for the quark-gluon 
plasma (QGP). The discussion of a phase diagram requires thermodynamic 
language. A phase transition from an initial color-deconfined QGP to a 
color-confined hadronic state (as it is supposed to occur in heavy-ion 
collisions) can only be reasonably well defined if the system under
study is in a state of approximate local thermodynamic equilibrium.
The application of thermal and hydrodynamic models to relativistic 
heavy-ion data is therefore more than a poor man's approach to
heavy-ion dynamics, it is rather a necessity for everybody who wants 
to convince himself and others that we succeeded in creating the 
quark-gluon plasma (QGP) and observing the phase transition accompanying 
its hadronization.

Of course, the models may fail; in fact, they must {\em necessarily} 
fail beyond a certain level of detail when applied to heavy-ion data. 
The reason is obvious: the collision systems are small, causing  
corrections to the infinite volume limit usually assumed in the 
thermodynamic approach, and they undergo a strong dynamical evolution on
time scales which are comparable to the microscopic thermalization time. 
Thermal models therefore can never provide more than a rough picture of 
the bulk of the phenomena, good for qualitative answers; on a more 
detailed and quantitative level, the failure of the thermal model will 
become manifest, and traces of genuine QCD {\em dynamics} (as opposed 
to {\em thermodynamics}) will show up. But when trying to assess 
{\em bulk} phenomena like QGP formation, we are not (in fact, we 
{\em must} not be) primarily interested in these deviations 
from thermodynamic behaviour and the traces of elementary QCD dynamics; 
the latter can be studied much more easily and cleanly in elementary 
lepton or hadron collisions. We should rather concentrate on the
rough global pattern of the data and try to understand them within a 
(nota bene: sufficiently sophisticated, see below) thermo- and 
hydrodynamic approach. On the other hand, if it turns out that not 
even the rough qualitative features of the data can be understood in 
this way and detailed hadronic dynamics is required even for a 
superficial understanding of the observations, then we should concede 
that our attempt to create ``hot QCD matter'' has failed.

Having argued in favor of a ``simple'' thermal approach to heavy-ion 
data, the next questions to be addressed are (i) the level of 
sophistication which the thermal model should have before being 
applicable to the description of particle production in nuclear 
collisions, (ii) which level of agreement between model and data
can at best be expected, and (iii) where to draw the line between 
agreement and disagreement when comparing model and data. This is 
what this contribution is about. At the end I will try to draw some 
conclusions about what we have learnt so far from the thermal 
analysis of heavy-ion data, and which further steps should be taken.

\section{Two types of ``thermal'' behaviour}
\label{sec2}

``Thermal'' behaviour can arise in {\em conceptually different}
ways, with different meanings of the ``temperature'' parameter $T$.
For us the two most important variants of ``thermal'' behaviour are the
following: 

{\bf 1.} The {\em statistical occupation of hadronic phase-space} with
minimum information. The latter is in practice provided by external 
constraints on the total available energy $E$, baryon number $B$, 
strangeness $S$ and, possibly, a constraint $\lambda_s$ on the overall 
fraction of strange hadrons. ``Thermal'' behaviour arises in this case
via the {\em Maximum Entropy Principle} in which the ``temperature'' $T$ 
and ``fugacities'' $e^{\mu_b/T}$, $e^{\mu_s/T}$ (which in the canonical
approach are replaced by so-called ``chemical factors''
\cite{becattini,BH97}) occur as Lagrange multipliers for the
constraints. Examples are nucleon emission from an evaporating
compound nucleus in low-energy nuclear physics and hadronization in
$e^+e^-$, $pp$ and $p\bar p$ collisions (hadron yields
\cite{becattini,BH97} and $m_\perp$-spectra~\cite{sollfrank}). The
number of parameters to fit the data in such a situation is equal 
to the number of ``conserved quantities'' (constraints), and it
reflects directly the information content of the fitted observable(s).
This type of ``thermal'' behaviour requires {\em no} rescattering and
{\em no} interactions among the hadrons, there is no isotropic pressure 
and no collective flow in the hadronic final state and, in fact, the 
concept of {\em local} equilibrium can {\em not} be applied. Of course, 
this type of ``thermal'' behaviour is not really what we are interested 
in in heavy ion collisions, except as a baseline against which to
differentiate interesting phenomena.

{\bf 2.} Thermalization of a non-equilibrium initial state by {\em
 kinetic equilibration} (rescattering). This does require (strong!)
interactions among the hadrons. Here one must differentiate between
{\em thermal} equilibration (reflected in the shape of the momentum
spectra), which defines the temperature $T$, and {\em chemical}
equilibration (reflected in the particle yields and ratios) which
defines the chemical potentials in a grand canonical description. The
first is driven by the {\em total} hadron-hadron cross section while
the second relies on usually much smaller {\em inelastic} cross
sections and thus happens more slowly. This type of equilibrium is
accompanied by pressure which drives collective flow (radial expansion
into the vacuum as well as directed and elliptic flow in non-central 
collisions). In heavy ion collisions it is realized {\em at most 
locally}, in the form of local thermal and/or chemical equilibrium -- 
due to the absence of confining walls there is never a global equilibrium. 
This is the type of ``thermal'' behaviour which we are searching for in 
heavy-ion collisions.   

I stress that {\em flow} is an unavoidable consequence of this type of
equilibration. Thermal fits without flow to hadron spectra are not
consistent with the kinetic thermalization hypothesis. Flow contains
information; it is described by three additional fit parameters
$\vec{v}(x)$. This information is related to the pressure history in
the early stages of the collision and thereby (somewhat indirectly) to
the equation of state of the hot matter. 

Most thermal fits work with global parameters $T$ and $\mu$ which, at
first sight, appears inconsistent with what I just said. Here the
role of freeze-out becomes important: freeze-out cuts off the
hydrodynamical evolution of the thermalized region via a kinetic
freeze-out criterium \cite{freeze} which involves the particle
densities, cross sections and the expansion rate. In practice freeze-out
may, but need not occur at nearly the same temperature everywhere 
\cite{freeze}. 

Clearly a thermal fit to hadron production data (if it works) is not
the end, but rather the beginning of our understanding. One
must still check the {\em dynamical consistency} of the fit
parameters $T_f$, $\mu_f$, $\vec{v}_f$: can one find equations of
state and initial conditions which yield such freeze-out parameters?
Which dynamical models can be excluded?

\section{The hadronic phase diagram}
\label{sec3}

\begin{figure}[t]
\caption[]{Compilation of freezeout points from SIS to SPS energies. 
Filled symbols: chemical freeze-out points from hadron abundances. Open
symbols: thermal freeze-out points from momentum spectra and
two-particle correlations. For each system, chemical and thermal
freeze-out were assumed to occur at the same value $\mu_B/T$.The shaded 
region indicates the parameter range of the expected transition to a QGP.
\label{figure1}}
\begin{indented}
\item[]\hspace{0cm}\epsfxsize 12cm \epsfbox{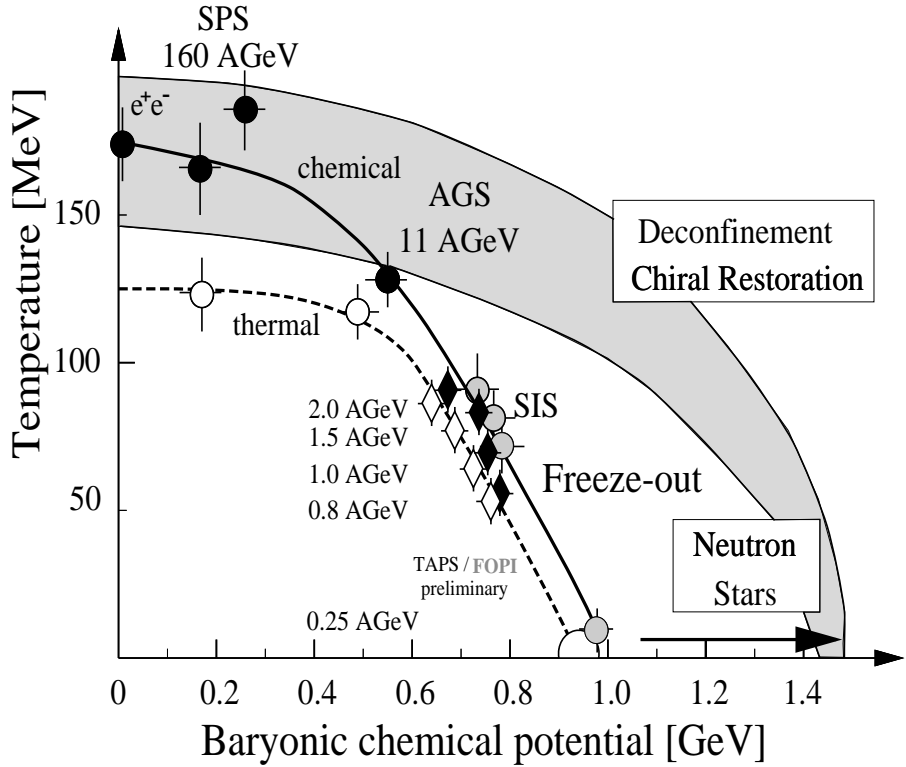}
\end{indented}
\end{figure}

In figure \ref{figure1} I show a recent version of the phase diagram 
for strongly interacting matter, with various sets of data points 
included \cite{S97,Hqm97,BMSqm97,Mqm97,CR98}. In the present section 
I discuss the meaning of this figure and explain how these data points 
were obtained. In the next section I will discuss some problems connected 
with the extraction procedures.

\subsection{Chemical freeze-out}
\label{sec3a}
\subsubsection{$e^+e^-$ and $pp$ collisions.}
\label{sec3a1}

Let me begin with the $e^+e^-$ data point in figure \ref{figure1}. (There 
is also a $pp$ point from Ref.~\cite{BH97} which was omitted for clarity.)
In spite of what I said about case {\bf 1.} above, a ``thermal''
analysis of hadron yields in elementary collisions \cite{becattini,BH97}
is still interesting. (Of course, in this case the canonical formalism
must be used, due to the small collision volume.) The interest arises 
{\em a posteriori} from the observed universality of the fit parameters, 
namely a universal ``hadronization'' or ``chemical freeze-out'' temperature 
$T_{\rm chem} = T_{\rm had} \approx 170$ MeV (numerically equal to the old
Hagedorn temperature $T_{\rm H}$ and consistent with the
inverse slope parameter of the $m_T$-spectra in $pp$ collisions
\cite{sollfrank}), and a universal strangeness fraction
$\lambda_s =$ $2\langle\bar s s\rangle/(\langle\bar u u \rangle +
\langle\bar d d\rangle)|_{\rm produced}$ $\approx 0.2{-}0.25$, almost 
independent of $\sqrt{s}$ \cite{becattini,BH97,BGS98}. 

This is most easily understood~\cite{BH97} in terms of a universal 
critical energy density $\epsilon_{\rm crit}$ for hadronization which, 
via the Maximum Entropy Principle, is parametrized by a universal 
``hadronization temperature'' $T_{\rm had}$ and which, according to 
Hagedorn, characterizes the upper limit of hadronic phase-space. 
Supporting evidence comes from the observed increase with $\sqrt{s}$ 
of the fitted fireball volume $V_f$ (which accomodates the increasing 
multiplicities and widths of the rapidity distributions). Although
higher collision energies result in larger {\em initial} energy
densities $\epsilon_0$, the collision zone subsequently undergoes more
(mostly longitudinal and not necessarily hydrodynamical) expansion until
$\epsilon_{\rm crit}$ is reached and hadron formation can proceed. The
systematics of the data can only be understood if hadron formation at
$\epsilon{>}\epsilon_{\rm crit}$ (i.e. $T{>}T_{\rm H}$ for the
corresponding Lagrange multipliers) is impossible. With this
interpretation, the chemical analysis of $e^+e^-$, $pp$ and $p\bar p$
collisions does provide one point in the $T$-$\mu_b$ phase diagram (see
figure \ref{figure1}). -- The only ``childhood memory'' of the collision 
system is reflected in the low value of $\lambda_s$, indicating suppressed 
strange quark production (relative to $u$ and $d$ quarks) in the early
pre-hadronic stages of the collision. 

\subsubsection{$AA$ collisions: strangeness enhancement.}
\label{sec3a2}

In this light the observation \cite{S97} of a chemical freeze-out
temperature $T_{\rm chem} \approx T_{\rm H} \approx 170$ MeV in heavy-ion 
collisions with sulphur beams at the SPS (figure \ref{figure1}), taken 
by itself, is not really interesting. It suggests that in heavy-ion
collisions hadronization occurs via the same statistical hadronic
phase-space occupation process as in $pp$ collisions. What {\em is}
interesting, however, is the observation \cite{BGS98,G96} that the global 
strangeness fraction $\lambda_s{\approx}0.4{-}0.45$ in $AA$ collisions 
is about a factor 2 larger than in $e^+e^-$ and $pp$ collisions. If $pp$ 
and S+$A$ collisions hadronize via the same mechanism, and in S+$A$ 
collisions the Maximum Entropy particle yields fixed at $T_{\rm had}$ 
are not modified by inelastic hadronic final state rescattering, this 
increase in $\lambda_s$ must reflect a {\em difference} in the properties 
of the {\em prehadronic} state! In nuclear collisions the prehadronic 
stage allows for more strangeness production, most likely due to a 
longer lifetime before hadronization. 

It was noted before \cite{BGS98,G96} that the global strangeness 
enhancement occurs already in collisions between medium size nuclei 
(S+S) and remains roughly unchanged in Pb+Pb collisions. In this 
conference we saw data from the WA97 collaboration \cite{WA97,Lietava,Evans} 
which provide two important further details: 

1. While the bulk of the strangeness enhancement from $p$+Pb to Pb+Pb 
collisions is carried by the kaons and hyperons ($\Lambda$, $\Sigma$), 
which are enhanced by about a factor 3 near midrapidity, the enhancement 
is much stronger for the doubly and triply strange baryons $\Xi$ and 
$\Omega$ and their antiparticles, with an enhancement factor of about 
17 (!) for $\Omega+\bar\Omega$ at midrapidity. The enhancement clearly 
scales with the strangeness content \cite{Lietava}, as naively expected 
in statistical and thermal models, but in stark contradiction to 
expectations based on the consideration of the respective production 
thresholds in hadronic (re)interactions.

2. In semicentral Pb+Pb collisions the enhancement grows linearly with 
the number of participating nucleons in the collision, so the enhanced 
yield of all measured strange hadron species per participating nucleon 
is {\em independent of the effective size of the colliding system} from 
about 150 to 400 participants \cite{WA97}. Where comparison is possible, 
this systematics even carries over to central S+S collisions \cite{Evans} 
with as few as 55-60 participating nucleons. So whatever causes the 
enhancement in Pb+Pb collisions (e.g. the existence of a color-deconfined 
prehadronic stage) is not particular to (semi)\-cen\-tral Pb+Pb collisions, 
but exists already in S+S collisions! I will return to the 
$A$-independence of this effect below.

At most half of the 100\% increase of global strangeness production 
in $AA$ collisions can be explained \cite{SBRS98} by the removal of 
canonical constraints in the small $e^+e^-$ and $pp$ collision 
volumes (which would be an interesting observation in itself 
because it would already imply that in nuclear collisions hadron 
production and the conservation of quantum numbers occurs no longer
on nucleonic, but indeed on nuclear length scales). The remainder of 
the increase must be due to extra strangeness production in the whole
fireball volume {\em before} hadronization. It is interesting to analyze  
in the same way the strong $\Omega+\bar\Omega$ enhancement in Pb+Pb 
(by a factor 17 relative to $p$+Pb \cite{Lietava}): of course, the $\Omega$ 
(carrying 3 units of strangeness) suffers a particularly strong canonical
suppression due to exact strangeness conservation in the small hadronization
volume of a $pp$ (or $p$+Pb) collision; for $T_{\rm had}\simeq 170$ MeV,
$\gamma_{\rm s}=0.5$ and $V=17.6$ fm$^3$ (as obtained by fitting $pp$
data at $\sqrt{s}=27$ GeV \cite{BH97}) $\Omega+\bar\Omega$ are 
canonically suppressed by a factor 12 relative to a grand canonical 
treatment \cite{Jocklpriv}. Again the observed enhancement effect in 
Pb+Pb collisions is considerably larger than expected from a simple 
removal of the canonical constraints.

The observed strangeness fraction $\lambda_s=0.45$ in nuclear collisions 
corresponds to a strangeness saturation coefficient \cite{R91} 
$\gamma_s\approx 0.7$ \cite{BGS98}. On the other hand, a value of 
$\gamma_s \approx 0.7$ in the hadronic final state may, in fact, be 
the upper limit reachable in heavy-ion collisions \cite{SBRS98} 
because the corresponding strangeness fraction agrees with that 
in a fully equilibrated QGP at $T_{\rm had} \approx 170$ MeV. If 
both strangeness and entropy are conserved or increase similarly 
during hadronization, $\gamma_s\approx 0.7$ in the Maximum Entropy 
particle yield after hadronization would be a universal consequence 
of a fully thermally and chemically equilibrated QGP before 
hadronization \cite{SBRS98}. The SPS data would then be completely
consistent with such a prehadronic state. 

The existence of a prehadronic stage without color confinement, both in 
S+S and Pb+Pb collisions at the SPS, is also suggested by an analysis of 
baryon/antibaryon ratios of different strangeness. This was stressed
at this conference by A. Bialas \cite{B98} who redid this analysis
with the new data following the ideas of Rafelski \cite{R91}.

According to figure \ref{figure1} chemical freeze-out in sulphur-induced 
collisions at the SPS appears to occur {\em right at the critical line}, 
i.e. immediately after hadronization. The SIS data, on the other hand, 
indicate much lower chemical freeze-out temperatures. The origin of this 
is probably due to longer lifetimes of the reaction zone at lower beam 
energies, allowing for some chemical re-equilibration by inelastic hadronic 
reactions. 

A tendency for some chemical re-equilibration after the hadronization of 
the proposed pre-hadronic stage may also be visible in the still 
preliminary Pb+Pb data at the SPS:  although thermal model analyses of 
these data still give wildly scattering results \cite{BGS98,BMSqm97,LR98}, 
some authors \cite{LR98} find chemical freeze-out temperatures in Pb+Pb 
below 140 MeV. A thermal model analysis of RQMD simulations also gives 
chemical freeze-out temperatures of 172 MeV in S+S, but of only 155 MeV 
in Pb+Pb collisions \cite{SHSX98}. Both analyses show a characteristic 
failure to reproduce the $\Omega$ and $\bar\Omega$ yields \cite{WA97}. 
This was interpreted \cite{SHSX98,HSX98} in terms of early freeze-out 
of these triply strange baryons due to their small interaction cross 
sections with other types of hadrons. It is interesting to observe that 
in the thermal analysis of the data \cite{LR98} the model {\em underpredicts} 
the measured $\Omega$ and $\bar\Omega$ yields (which prefer a higher 
freeze-out temperature $T\geq 170$ MeV \cite{BMSqm97,BGS98}) while in 
RQMD, which is known to produce too few $\Omega$ and $\bar \Omega$ 
baryons in $pp$ and $pA$ collisions, the thermal model {\em overpredicts}
the simulated yields. All this points to the $\Omega$ and $\bar\Omega$ as 
relatively early hadronic messengers, and the message they seem to carry 
in the Pb+Pb {\em data} is again that of the existence of a prehadronic 
stage with enhanced global strangeness which hadronizes statistically at 
$T_{\rm had} = T_{\rm H} \simeq 170$~MeV. 

\subsection{Flow and thermal freeze-out}
\label{sec3b}
 
The other important observation in the hadronic sector of nuclear
collisions is that of collective flow (radial expansion flow, directed
and elliptical flow). It is usually extracted from the shape of the
single-particle momentum distributions. Radial flow, for example,
leads to a flattening of the $m_\perp$-spectra. For the analysis one
must distinguish two domains. In the relativistic domain
$p_\perp{\gg}m_0$ the inverse slope $T_{\rm app}$ of all particle
species is the same and given by the blueshift formula \cite{freeze}
$T_{\rm app}{=}T_f \sqrt{(1{+}\langle v_\perp\rangle)/(1{-}\langle
  v_\perp\rangle)}$. This formula does not allow to disentangle the
average radial flow velocity $\langle v_\perp\rangle$ and freeze-out
temperature $T_f$. In the non-relativistic domain $p_\perp{\ll}m_0$
the inverse slope is given approximately by $T_{\rm app}{=}T_f{+}m_0
\langle v_\perp^2 \rangle$, and the rest mass dependence of the
``apparent temperature'' (inverse slope) allows to determine $T_f$ and
$\langle v_\perp^2 \rangle$ separately. (In $pp$ collisions no
$m_0$-dependence of $T_{\rm app}$ is seen \cite{NuXu}.) Plots of
$T_{\rm app}$ against $m_0$ were shown in several talks at this
conference, showing that the data follow very nicely this systematics,
from SIS to SPS energies.

A notable exception are the $\Omega$-spectra of WA97 \cite{WA97}
which are steeper than expected from this formula. Again, as in the 
above discussion of their abundance, this reflects their character
as ``early hadronic messengers'' \cite{HSX98}: the $\Omega$ and 
$\bar \Omega$ are the only baryons which (due to quantum number mismatch) 
do not have a strong resonance with pions, the most abundant particles 
in the fireball. Since resonance scattering is the most efficient 
thermalization mechanism in a dense hadronic system, the $\Omega$ and 
$\bar \Omega$ momentum distributions freeze out earlier than those of 
all other baryons. This implies \cite{HSX98} that they cannot efficiently 
pick up the collective transverse flow which builds up among the pions 
in the later stages of the expansion, and their spectra reflect the much 
weaker collective transverse flow in the early collision stages, 
just after hadron formation.

[This also illustrates the important role of the baryon contamination 
in the hot fireball: in Pb+Pb collisions at the SPS the pion and baryon 
spectra decouple late and cool down to rather low temperatures of about 
120 MeV (see below) because the pions are ``glued'' together by the 
baryons via resonance scattering. At RHIC this glue will be less efficient 
since near midrapidity there will essentially exist only baryon-antibaryon 
pairs at rather low thermal equilibrium abundances. It is thus expected 
that at RHIC thermal decoupling occurs at somewhat higher freeze-out 
temperatures, closer to the hadronization phase transition.] 

The separation of collective flow and random thermal motion from an 
analysis of single particle spectra is not uncontroversial. The main
reason is that the fitted values for $T$ and $v_\perp$ tend to be 
strongly correlated. To break the correlation one must study spectra
of hadrons with different masses in the low-$p_\perp$ region which,
on the other hand, is contaminated by post-freeze-out resonance decays.
As a consequence, fits done in different $p_\perp$ windows tend to give
different results. 

\begin{figure}[t]
\caption[]{Thermal freeze-out temperature and transverse flow 
velocities extracted from fits to the transverse momentum spectra
of negative hadrons ($h^-$) and deuterons (d) and to the transverse
HBT radius ($2\pi$-BE). The shaded area indicates the overlap region
near $T_{\rm f.o.} \approx 120$ MeV and $v_\perp \approx 0.55$ $c$.
(From Ref.~\protect\cite{NA49HBT}.)
\label{figure2}}
\begin{indented}
\item[]\hspace{0cm}\epsfxsize 12cm \epsfbox{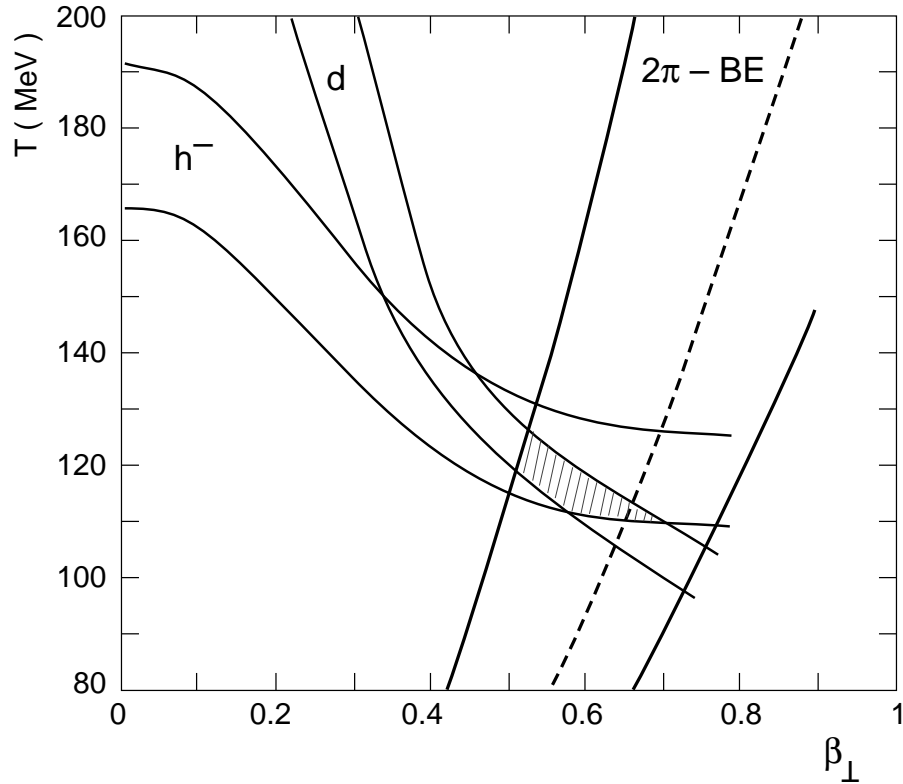}
\end{indented}
\end{figure}

A clearer determination of the transverse collective flow comes 
from a direct measurement of the flow-induced space-momentum
correlations via the $M_\perp$-dependence of the two-pion HBT radii
\cite{H96}. As shown in \cite{NA49HBT,Wiedemann} the correlations between
temperature and transverse flow in a fit of the single particle transverse 
momentum spectra and of the transverse two-particle HBT radii are essentially
orthogonal to each other (see figure \ref{figure2}), and the combined
analysis of spectra and correlations allows for a clean separation of 
random thermal motion from collective flow. 

This analysis \cite{NA49HBT,Wiedemann} gave rise to the open circle for 
the SPS data in figure \ref{figure1}, indicating the point of thermal
decoupling in 158 $A$ GeV/$c$ Pb+Pb collisions ($T_{\rm therm} \approx
120$ MeV, $\langle v_\perp \rangle \approx 0.5\, c$). It is consistent 
with a comparison of the spectral slopes for different mass hadrons 
by K\"ampfer \cite{kaempfer}. Earlier analyses of S+S data \cite{SSH93} 
showed a thermal decoupling at $T_{\rm therm} \approx 140-150$ MeV, 
$\langle v_\perp \rangle \approx 0.25-0.35\, c$. The open circle for the 
AGS as well as the open quadrangles for the SIS in figure \ref{figure1} 
were obtained similarly as in \cite{kaempfer,SSH93} by comparing 
transverse momentum spectra of particles with different masses 
(see \cite{BMSqm97,Mqm97} for references). All open symbols correspond 
to heavy collision systems (Pb+Pb, Au+Au). The dashed line connects them 
by eye in an attempt to construct a ``thermal freeze-out curve'' for 
heavy-ion collisions of size 200+200.
 
It is interesting to analyze the $A$-dependence of the various hadronic 
observables, i.e. of strangeness enhancement, chemical and thermal 
freeze-out temperatures and radial transverse flow. This discussion should 
include the available information on the impact parameter dependence 
in Pb+Pb collisions since collisions at different impact parameters also
involve different numbers $N_{\rm part}$ of participating nucleons. 
Whereas the freeze-out temperatures (certainly the thermal freeze-out 
temperature, but perhaps also the chemical one, see above) seem to come 
down with increasing size of the collision system, while at the same time 
the strength of the radial transverse flow goes up, the strangeness 
enhancement (i.e the strange particle production per participating 
nucleon relative to $pp$ and $pA$ collisions) appears to be independent 
of the number of $N_{\rm part}$, at least in the range $60 \leq 
N_{\rm part} \leq 400$. The buildup of radial flow is largely (although
not exclusively) a hadronic reinteraction phenomenon \cite{HSX98}; the
same is true for the freeze-out process. The $N_{\rm part}$-dependence 
of both features can be explained in terms of the longer lifetime of the
reaction zone as $N_{\rm part}$ increases, giving the system more time
to re-equilibrate, cool down and develop collectivity. 

In contrast, the $A$-independence of the strangeness enhancement 
features suggest that they are not due to hadronic re-interactions, 
but originate in a prehadronic phase with properties which are 
essentially independent of the system size once $N_{\rm part} \geq 50$ 
or so. My interpretation of these facts is that at SPS energies the 
energy density threshold for QGP has been overcome by a sizeable margin 
\cite{Rio97}, and that even in small collision systems a deconfined phase 
is created which interacts sufficiently long and sufficiently strongly 
to approximately saturate strangeness production. One can even argue 
that isotropic pressure (a signature of local thermalization) must be 
present at this early stage \cite{Rio97}, and that the observed elliptic 
flow in non-central Pb+Pb collisions \cite{NA49flow} actually signals 
this early pressure \cite{private}. A future systematic investigation 
also of the range $1 < N_{\rm part} < 100$ would be very useful to 
study the onset and saturation of thermal and collective behaviour 
as the size {\em and lifetime} of the collision system increases.
It will {\em not}, however, be an efficient method to study the onset of 
deconfinement -- for that one must go to {\em lower} beam energies. 

\section{Limitations of thermal model analyses}
\label{sec4}

After having explained how the data points in figure \ref{figure1} were 
obtained, I would now like to ask the notorious David Mermin question 
\cite{Mermin} ``What's wrong with this phase diagram?'' In other words, 
I want to point out in more detail certain unavoidable problems with 
thermal model analyses of heavy-ion data. Only by remaining conscious 
of the limitations of the thermal approach and avoiding the 
overinterpretation of uncontrollable details one can fully exploit
its power in providing essential qualitative insight into the physics
of heavy-ion collisions.

\subsection{Rapidity dependence of particle ratios}
\label{sec4a}

The first problem is of purely practical nature: no heavy-ion experiment 
so far has full $4\pi$ acceptance for identified particles, and particle
spectra are available in restricted windows of $p_\perp$ and $y$ only.
The observed nearly exponential form of the $p_\perp$-spectra allows for 
an extrapolation of the yields to the full transverse momentum range
without introducing large uncertainties (at least if the acceptance 
covers sufficiently low values of $p_\perp$). A similar extrapolation 
of the rapidity spectra is not possible: even in a system which is in
perfect local thermal equilibrium, particles with different masses 
tend to have strongly different rapidity distributions. In practice
these differences in the shape of the rapidity spectra are even stronger, 
in the sense that even particles and antiparticles (with obviously identical 
masses) have different rapidity distributions. Thus there is essentially
no way of extrapolating the rapidity distributions without really measuring
them.

That this is a serious problem for thermal model analyses is illustrated
by the following example: consider a stationary, spherical fireball in
global thermodynamic equilibrium. It emits hadrons with the following
rapidity distributions:
 \begin{equation}
 \label{rapid}
   {dN_i\over dy} \sim e^{-m_i\cosh y/T} \left[ 1 + 2 {T\over m_i \cosh y}
   + 2 \left( {T\over m_i \cosh y} \right)^2\right]
 \end{equation}
These resemble Gaussians with widths $\Gamma_i\approx 2.35\sqrt{T\over m_i}$ 
(the approximation being valid for $m_i \gg T$). Clearly, the particle 
ratios $(dN_i/dy)/(dN_j/dy)$ then depend strongly on the position of 
the rapidity interval $dy$: away from $y=0$ heavy particles will be 
much more strongly suppressed relative to light particles than near 
$y=0$. In this case a measurement of particle yields in a small 
rapidity window is completely useless for a thermal model analysis 
(irrespective of whether the window is located near $y=0$ or at $y\ne 0$) 
unless it is {\em known a priori} that the radiator is a stationary
spherical fireball. 

The presence of (strong) longitudinal flow in relativistic heavy-ion 
collisions does not help very much in this connection; only in the limit
of infinite beam energy with exact longitudinal boost-invariance due to 
Bjorken scaling, resulting in flat rapidity distributions, is it possible
to base a thermal analysis on data in a finite, narrow rapidity interval.
At SPS energies and below, where Bjorken scaling is not observed, thermal 
fits of data in finite rapidity windows require that the thermal model yields
are cut to the actual experimental acceptance; this induces serious 
dependences on the detailed model assumptions, for example about the 
strength and profile of the longitudinal and transverse flow of the source. 
This model-dependence was recently studied in some detail in 
Ref.~\cite{soll98}.

Flow effects drop largely out, however, if one works with particle 
ratios obtained from $4\pi$ yields. (Please note that this requires 
a {\em measurement} of some sort, not a blind extrapolation of data 
from a small window in $y$ and $p_\perp$ to full momentum space!) The 
insensitivity to hydrodynamic flow becomes exact if the freeze-out 
temperature and chemical potential is everywhere the same. If freeze-out 
occurs on a sharp hypersurface $\Sigma$, the total yield of particle 
species $i$ is then given by
 \begin{equation}
 \label{flow}
   N_i = \int {d^3p\over E} \int_\Sigma p^\mu d^3\sigma_\mu(x)\, f_i(x,p)
       = \int_\Sigma d^3\sigma_\mu(x) \, j_i^\mu(x)\,,
 \end{equation}
where
 \begin{equation}
 \label{current}
   j_i^\mu(x) = \int d^4p\, 2\theta(p^0)\delta(p^2-m_i^2)
   \, p^\mu {g_i\over e^{[p{\cdot}u(x) - \mu_i]/T} \pm 1}
 \end{equation}
is the number current density of particle species $i$. In thermal 
equilibrium it is given by $j_i^\mu(x)=\rho_i(x)\, u^\mu(x)$ with
 \begin{eqnarray}
 \label{dens}
   \rho_i(x) &=& u_\mu(x) j_i^\mu(x) 
     = \int d^4p\, 2\theta(p^0)\delta(p^2-m_i^2)\, p{\cdot}u(x) \, 
       f_i\bigl(p{\cdot}u(x);T,\mu_i\bigr)
 \nonumber\\
   &=& \int d^3p'\, f_i(E_{p'};T,\mu_i) = \rho_i(T,\mu_i).
 \end{eqnarray}
Here $E_{p'}$ is the energy in the local rest frame at point $x$.
The total particle yield of species $i$ is therefore
 \begin{equation}
 \label{yield}
   N_i = \rho_i(T,\mu_i) \int_\Sigma d^3\sigma_\mu(x) u^\mu(x) 
       = \rho_i(T,\mu_i)\, V_\Sigma(u^\mu)
 \end{equation}
where only the total comoving volume $V_\Sigma$ of the freeze-out 
hypersurface $\Sigma$ depends on the flow profile $u^\mu$. Thus the 
flow pattern drops out from ratios of $4\pi$ yields which therefore 
depend only on $T$ and the chemical potentials. These considerations 
are easily generalized to ``fuzzy freeze-out'' (i.e. freeze-out from 
a space-time volume rather than from a sharp hypersurface): as long as 
$T$ and $\mu_i$ are the same everywhere, $4\pi$ particle ratios are 
independent of the collective dynamics of the source.

For heavy-ion collisions at SPS energies and below one should therefore
perform a thermal analysis on $4\pi$-integrated yields and not on
particle ratios inside small rapidity windows. This requires a strong 
experimental effort to measure the rapidity distributions of as many particle
species as possible over the full rapidity range.

\subsection{Non-constant thermodynamic parameters at freeze-out}
\label{sec4b}

The second, even more serious problem is the observation that in reality
freeze-out {\em does not happen} at constant temperature and chemical 
potential. For example, it was shown in Ref.~\cite{Slotta95} that a 
successful thermal description of the rapidity distributions of hadrons
created in 200 $A$ GeV S+S collisions (in particular the different
shapes of the rapidity distributions for $\Lambda$ and $\bar \Lambda$, $K^+$
and $K^-$) requires not only strong longitudinal flow, but also a baryon 
chemical potential $\mu_i(\eta)$ which depends on the longitudinal 
position in the fireball: the central rapidity region is baryon-poorer 
than the target and projectile fragmentation regions. A second example 
demonstrating this type of problem is the observed rapidity dependence of 
the $p_\perp$-slopes of the $h^-$ and proton spectra \cite{Jacobs} and of 
the $K_\perp$-slopes of the transverse HBT radius $R_\perp$ 
\cite{NA49HBT,Schoenf}. According to a simultaneous analysis (as 
discussed in section \ref{sec3b} above) of spectra and correlations from
Pb+Pb collisions at the SPS by Sch\"onfelder \cite{Schoenf} the 
decrease of the inverse slope parameters away from midrapidity must 
be attributed to {\em both} a reduction of the transverse collective 
flow {\em and} of the freeze-out temperature $T(\eta)$. If true, this 
would speak against a constant freeze-out temperature.

In such a situation the thermal fit replaces the functions $T(\eta)$ and
$\mu_i(\eta)$ by suitably averaged values $\bar T$, $\bar\mu_i$ (see 
Ref.~\cite{soll98} for a recent detailed study). Obviously the fit will
then not be perfect: particle yields from a system in perfect local 
thermodynamic equilibrium (as, e.g., assumed in all hydrodynamic 
simulations), but with {\em spatially varying} temperature and chemical 
potentials, cannot be exactly recovered by a fit with {\em constant}
temperature and chemical potential. In practice, this does not appear to
be a very serious problem if one believes that the results from a recent 
thermal model analysis of particle yields from a hydrodynamic simulation
\cite{soll98} are representative for realistic situations: the freeze-out
temperature reconstructed from the fit nearly coincided with the input
temperature at which the hydrodynamic evolution was stopped, and the fitted
chemical potentials agreed approximately with their average values across 
the freeze-out surface. Nevertheless, small differences remain between
the real yields from the hydrodynamic simulation and the yields returned
by the thermal model fit. In a least mean square fit these {\em systematic} 
deviations would lead to a value for $\chi^2$/d.o.f. which increases above 
all limits as the statistical error of the simulated (``measured'')
yields is further and further reduced, even though the system was, by 
construction, in perfect local thermal equilibrium.

This illustrates that global thermal fits to heavy-ion data can never 
be fully successful, due to the dynamics of the collision and its 
intricate influence on the freeze-out process. For this reason one must 
not expect too much from the thermal model -- a perfect fit with extremely 
small $\chi^2$/d.o.f. is not necessarily a good and often rather a bad
sign, indicating accidental error correlations, e.g. due to the use 
of redundant fit parameters.

This leaves us with the question where to draw the line between ``good'' 
and ``bad'' thermal model fits, between success and failure of a 
thermodynamic description of relativistic heavy-ion collisions. The above
discussion should have made it clear that $\chi^2$/d.o.f. is not a good
criterium for answering this question. On the other hand, a fit which 
reproduces particle yields which cover a range of more than three orders 
of magnitude with individual deviations of less than $\pm$25\% \cite{BH97} 
is obviously not bad. Quantitative model studies like those presented 
in \cite{soll98} give us guidance for separating the grain from the 
straw; when supplemented by a thermal analysis of microscopic kinetic 
simulations as those presented at this meeting by Larissa Bravina 
\cite{Bravina}, they are the foundations which we can use when 
collecting arguments in favor or against the formation of thermalized 
hot hadronic matter and quark-gluon plasma.

\section{Conclusions}
\label{sec5}

Let me summarize shortly: a thermal + flow analysis of yields, spectra 
and 2-particle correlations in S+$A$ and Pb+Pb collisions at the CERN 
SPS suggests

\begin{itemize}
 
\item the formation of a {\em prehadronic} state in which twice as 
      much strangeness is produced as in $pp$ and $pA$ collisions and
      in which quarks are uncorrelated, i.e. they are not bound into 
      color singlets;

\item statistical hadronization of this state at $T_{\rm had} 
      \approx 170 \pm 20$ MeV with hadron abundances controlled 
      by the {\em Maximum Entropy Principle};

\item rapid decoupling of the particle abundances, with chemical freeze-out
      temperatures $T_{\rm chem} \approx T_{\rm had}$ in sulphur-induced
      and $T_{\rm chem} \leq T_{\rm had}$ in Pb+Pb collisions;

\item elastic rescattering among the hadrons (dominated by $s$-channel
      resonances) after hadronization from the prehadronic state which
      leads to a cooling of the momentum spectra and the generation of 
      (more) collective flow; 

\item finally, thermal freeze-out at $T_{\rm therm} \approx 140-150$ MeV
      in S+$A$ and at $T_{\rm therm} \approx 120\pm 10$ MeV in Pb+Pb
      collisions, with average transverse collective flow velocities of
      order $\langle v_\perp \rangle \approx 0.4-0.5\,c$.  

\end{itemize}

Smaller collision systems distinguish themselves from larger ones not 
primarily by the achieved maximal energy density, but by the occupied 
collision volume in space and time. Compared to S+S collisions, Pb+Pb 
collisions live longer until thermal freeze-out, expand more in the 
tranverse direction, develop more transverse collective flow and cool 
down to lower (chemical and thermal) freeze-out temperatures. Of course,
this view disagrees with the (present) majority opinion (I refer to the 
respective contributions to the Proccedings of ``Quark Matter '97'' 
\cite{QM97}) that the critical energy density for deconfinement can 
be crossed at fixed beam energy of 160 $A$ GeV by changing the size 
of the projectile and target or the collision centrality, and that
the ``anomalous'' $J/\psi$ suppression observed in central Pb+Pb 
collisions as a function of produced transverse energy \cite{QM97,NA50}
signals this transition. 

I find this interpretation irreconcilable with the systematics of 
light hadron production as discussed in this talk; a more consistent 
interpretation rests on the observation \cite{NA50} that, as one 
increases $N_{\rm part}$, one first sees ``anomalous'' suppression of 
the weakly bound $\psi'$, then (in semiperipheral Pb+Pb collisions) 
the ``anomalous'' suppression of the more strongly bound $\chi_c$ states
(indirectly, via the disappearance of their 32\% feed-down contribution 
to the measured $J/\psi$ yield), and only in very central Pb+Pb 
collisions the disappearance of directly produced $J/\psi$'s which 
are {\em very} strongly bound (this last part of the supression pattern 
still remains to be confirmed by an improved measurement at very high 
$E_T$). This suggests to me that what NA50 is seeing is {\em not the 
onset of deconfinement} (the latter is there even in S+$A$ collisions), 
but the dissociation of more and more strongly bound heavy quark states 
(respectively the removal of the corresponding components in the 
$c\bar c$ wavefunction) by collisions with the dense partonic medium 
in the early stages of the collision. For the more strongly bound 
states most of these collisions will be subthreshold; for this reason
a longer lifetime of the dense early stage, which is achieved in larger
collision systems or more central collisions, is crucial for an efficient
destruction not only of the weakly bound $\psi'$, but also of the more
strongly bound $\chi_c$ and $J/\psi$. Charmonium suppression is thus, in 
my opinion, more a {\em lifetime effect} than a {\em deconfinement signal}
(although the necessary high density of scatterers with sufficiently large
cross sections probably requires deconfinement, too).

The picture which I have painted is, I believe, intrinsically consistent.
It is sufficiently simple to be attractive but also sufficiently 
sophisticated not to be unrealistic. It may not be unique, not least
because of the intrinsic systematic uncertainties associated with thermal 
model analyses which I pointed out and not all of which are quantitatively 
understood. What is urgently needed is more high-quality data on the
chemistry of Pb+Pb collisions, the reconciliation of some puzzling
discrepancies between different experiments as discussed at this meeting,
and an improved systematics of the impact parameter and $A$-dependences,
both in the light and heavy hadron sector.  

\ack
The author acknowledges the hospitality at the INT (Seattle) in March/April 
1998 where discussions with many colleagues allowed him to sharpen
the arguments presented here. This work was supported by BMBF, DFG
and GSI.


\section*{References}
\begin{thebibliography}{99}

\bibitem{becattini}
  Becattini F 1996 \ZP C {\bf 69} 485;
  Becattini F 1999 \JPG {\bf 25}, in press.
\bibitem{BH97}
  Becattini F and Heinz U 1997 \ZP C {\bf 76} 269
\bibitem{sollfrank}
  Sollfrank J, Koch P and Heinz U 1991 \ZP C {\bf 52} 593
\bibitem{freeze}
  Schnedermann E, Sollfrank J and Heinz U 1993 {\it Particle
    Production in Highly Excited Matter} NATO ASI Series B{\bf 303} 
    (New York: Plenum) p~175
\bibitem{S97}
  For a recent compilation of theoretical analyses see Sollfrank J 1997 
  \JPG {\bf 23} 1903, especially figure 5 in the electronic preprint 
  nucl-th/9707020 which was not included in the published version. 
\bibitem{Hqm97}
  Heinz U 1998 \NP A {\bf 638} 357c
\bibitem{BMSqm97}
  Braun-Munzinger P and Stachel J 1998 \NP A {\bf 638} 3c
\bibitem{Mqm97}
  Metag V 1998 \NP A {\bf 638} 45c
\bibitem{CR98}
  Cleymans J and Redlich K  1998 \PRL {\bf 81} 5284
\bibitem{BGS98}
  Becattini F, Ga\'zdzicki M and Sollfrank J 1998 {\it Eur. Phys. J.} 
  C {\bf 5} 143
\bibitem{G96}
  Ga\'zdzicki M 1996 {\it APH N.S.: Heavy Ion Physics} {\bf 4} 33, and
  references therein
\bibitem{WA97}
  Andersen E \etal (WA97 Coll.) 1998 \PL B {\bf 433} 209\\
  Kr\'alik I \etal (WA97 Coll.) 1998 \NP A {\bf 638} 115c\\
  Holme H K \etal (WA97 Coll.) 1997 \JPG {\bf 23} 1851
\bibitem{Lietava}
  Lietava R \etal (WA97 Coll.) 1999 \JPG {\bf 25}, in press;
  Caliandro R \etal (WA97 Coll.) 1999 \JPG {\bf 25}, in press
\bibitem{Evans}
  Evans D \etal (WA85/WA94 Coll.) 1999 \JPG {\bf 25}, in press.
\bibitem{SBRS98}
  Sollfrank J, Becattini F, Redlich K and Satz H 1998 \NP A {\bf 638} 399c
\bibitem{Jocklpriv}
  Sollfrank J 1998 private communication
\bibitem{R91}
  Rafelski J 1991 \PL B {\bf 262} 333
\bibitem{B98}
  Bialas A 1998 \PL B in press (hep-ph/9808434)
\bibitem{LR98}
  Letessier J and Rafelski J 1998 hep-ph/9807346
\bibitem{SHSX98}
  Sollfrank J, Heinz U, Sorge H and Xu N 1999 \JPG {\bf 25}, in press
\bibitem{HSX98}
  van Hecke H, Sorge H and Xu N 1998 \PRL {\bf 81} 5764
\bibitem{NuXu}
  Xu N \etal (NA44 Coll.) 1996 \NP A {\bf 610} 175c
\bibitem{H96}
  Heinz U 1996 \NP A {\bf 610} 264c and references therein
\bibitem{NA49HBT}
  Appelsh\"auser H \etal (NA49 Coll.) 1998 {\it Eur. Phys. J.} C {\bf 2} 661
\bibitem{Wiedemann}
  Wiedemann U A, Tom\'a\v sik B and Heinz U 1998 \NP A {\bf 638} 475c
\bibitem{kaempfer}
  K\"ampfer B \etal 1997 \JPG {\bf 23} 2001
\bibitem{SSH93}
  Schnedermann E, Sollfrank J and Heinz U 1993 \PR C {\bf 48} 2462\\
  Schnedermann E and Heinz U 1994 \PR C {\bf 50} 1675
\bibitem{Rio97}
  Heinz U 1998 in: {\it Relativistic Aspects of Nuclear Physics} \
  (Proceedings of the Fifth Rio de Janeiro International Workshop, 
  T. Kodama \etal, eds.) (Singapore: World Scientific) p~19
\bibitem{NA49flow}
  Appelsh\"auser H \etal (NA49 Coll.) 1998 \PRL {\bf 80} 4136
\bibitem{private}
  Poskanzer A and Sorge H 1998 private communications and work in preparation
\bibitem{Mermin}
  Mermin D 2000 {\it Physics Today} {\bf 52} to be published
\bibitem{soll98}
  Sollfrank J 1998 Preprint nucl-th/9811078
\bibitem{Slotta95}
  Slotta C, Sollfrank J and Heinz U 1995 {\it Strangeness in 
  Hadronic Matter}, AIP Conference Proceedings {\bf 340} 
  (Woodbury: AIP Press) p~462
\bibitem{Jacobs}
  Jacobs P 1997 {\it Proceedings of ICPA-QGP97}, Jaipur, India, March 17-21, 
  1997, NA49 note number 123 
\bibitem{Schoenf}
  Sch\"onfelder S 1997 PhD thesis, MPI f\"ur Physik, M\"unchen, NA49 
  note number 143
\bibitem{Bravina}
  Bravina L V \etal 1999 \JPG {\bf 25}, in press
\bibitem{QM97}
  Hatsuda T \etal (eds.) 1998 \NP A {\bf 638}
\bibitem{NA50}
  Abreu M C \etal (NA50 Coll.) 1997 \PL B {\bf 410} 337\\
  Romana A \etal (NA50 Coll.) 1998 {\it XXXIIIrd Rencontres de Moriond}, 
  21-28 March 1998, Les Arcs, France
\end{thebibliography}
\end{document}